# Survey of Recent Multi-Agent Reinforcement Learning Algorithms Utilizing Centralized Training

Piyush K. Sharma[a], Rolando Fernandez[a], Erin Zaroukian[a], Michael Dorothy[a], Anjon Basak[b], and Derrik E. Asher[a]

[a]US DEVCOM Army Research Laboratory, Adelphi, MD 20783
[b]Oak Ridge Associated Universities, 4692 Millennium Drive, Suite 101, Belcamp, MD 21017

**ABSTRACT**

Much work has been dedicated to the exploration of Multi-Agent Reinforcement Learning (MARL) paradigms implementing a *centralized learning with decentralized execution* (CLDE) approach to achieve human-like collab- oration in cooperative tasks. Here, we discuss variations of centralized training and describe a recent survey of algorithmic approaches. The goal is to explore how different implementations of information sharing mechanism in centralized learning may give rise to distinct group coordinated behaviors in multi-agent systems performing cooperative tasks.

## 1. INTRODUCTION

Reinforcement learning (RL) is a technique that can be used to explore decision-making and is modeled as a *Markov Decision Process (MDP)*. RL typically computes the optimal policy by taking the best action at each state to maximize reward over time.[1, 2] The underlying mathematical framework usually requires iterating over the *Bellman equations* and using dynamic programming.[3] Conventional single agent RL has long been studied where an agent interacts with an environment by taking actions from a current state and measuring the effect of those actions by received rewards, then transitioning to a new state (Figure 1).

### 1.1 Multi-Agent Reinforcement Learning

*A multi-agent reinforcement learning (MARL)* frame- work is fundamentally no different from single agent RL. In fact, a simple approach is to independently train all agents. In this MARL approach, each agent considers all the other agents to be part of the envi-ronment, and each agent must learn its own policy. This example illustrates a *Fully Decentralized* learning approach.[4–6] An alternative simple MARL approach is the implementation of *Centralized Learning with De- centralized Execution (CLDE)*. An example of this type of approach is to take in the state of the environment and return an action for each agent in the form of single joint action vector, thus learning a single policy for all agents. The primary focus of this paper is to explore the centralized learning mechanism in MARL algorithms which is covered in the majority of this paper.

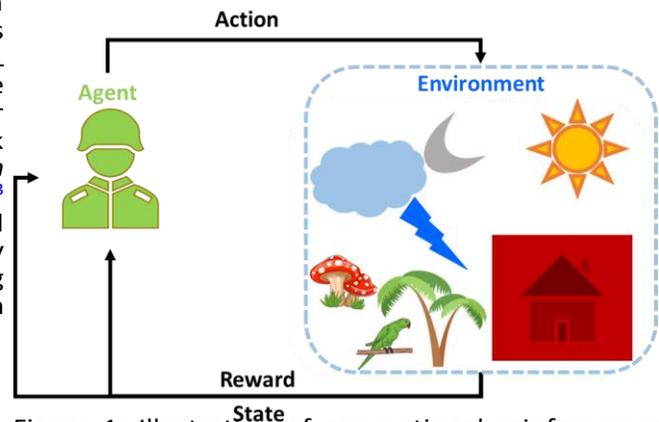

Figure 1: Illustration of conventional reinforcement learning. At each step, agent interacts with the environment by taking policy based action, receives a reward and transitions to a new state.

---

Corresponding author (Piyush K. Sharma):

CLDE is an approach that has been shown to significantly improve the performance of *Multi-Agent Systems(MAS)*[7] when compared to a fully decentralized training algorithm. The core concept behind these CLDE algorithms is to provide more complete state information to agents during training, such that their individual experiences are aggregated in some form (i.e., centralization of training), that allow agents to produce coordinated actions in testing (i.e., after training is completed or learning is turned off) from incomplete local observations (partial observed states).

Such artificial intelligence (AI) systems utilize techniques that enable agents to learn by interacting with their environments and each other in action-observation loops, or in other words, experiences. Given enough experiences, these MAS tend to adopt coordinated behaviors in cooperative tasks that can resemble those expected from teams of humans performing the same tasks.[8] Therefore, in such a scenario, a centralized critic (value function to measure actions) will have complete state information, while decentralized actors (policy function to control agent behavior) will act based upon local partial state observations. A detailed discussion on *actor-critic* methods is provided in Subsections 2.1 and 2.2.

In a given environment, a policy-based RL algorithm assigns a random initial policy to find a value function of that policy. At each successive step, this function is used to find the updated policy, and the process is repeated until the RL algorithm finds an optimal policy. On the other hand, in a value-based RL algorithm, this is done by selecting a random initial value function which is updated recursively until an optimal value function is found. Therefore, a policy corresponding to the optimal value function is an optimal policy which is implicitly updated using the value function. A value-based algorithm, such as Deep Q-Learning, may provide a holistic measure of an agent's action.[9] Our theoretical framework sheds light on an individual agent's actions and missing or asymmetric information about the agent's actions/attributes limiting their chances to collaborate.

In this work we explore recent methods and algorithmic approaches that have been used to train multi-agent systems to perform cooperative tasks with different implementations of centralized training in their information sharing mechanism (e.g., reward, gradient, action, parameter, observation/state space sharing). We discuss robustness challenges arising from an adversarial agent in the aforementioned *CLDE* settings. We also discuss the implications and expectations of providing a classification of all MARL algorithmic approaches and categorize them on the basis of their information sharing mechanism (e.g., reward, gradient, action, parameter, observation/state space sharing) on cooperative behavior in multi-agent systems. This survey of recent centralized learning approaches in MARL is meant to serve as a comprehensive summary and not an exhaustive review of other (decentralized/distributed) learning approaches. We also discuss how different forms of centralized learn- ing can overcome the *credit assignment problem*,[10] a central issue in reinforcement learning, that leads to poor convergence properties. We summarize MARL models and algorithms with their limitations and scientific gaps that need to be filled in novel CLDE approaches for multi-agent systems. Our study provides a comprehensive overview of the most recent advances in various types of reward strategies adopted by these approaches, and their importance in sequential decision-making with MARL algorithms.

In the remaining Subsections 1.2 and 1.3, we discuss the various problems faced by MAS, and their importance in the Military's *Multi Domain operations (MDO)* respectively. In Section 2, We discuss actor-critics method and how A2C and A3C based implementations can improve it. Our discussion extends to different types of learning approaches (value, policy, and hybrid) in MARL techniques. In Section 3 we introduce recent MARL algorithms and discuss their challenges in-depth, such as the *credit assignment* problem, the *lazy agents* problem related the to partial observability, and *heterogeneity* arising from the different learning algorithms. We extend this discussion by summarizing differences in centralized learning approaches and highlighting information shared during centralized learning. In Section 4, we draw conclusion based on the summary of this work.

### 1.2 Challenges in Multi-Agent Systems

In MARL the reward associated with a specific state can vary over time, and as the number of agents increase the non-stationarity problem suffers from the curse of dimensionality.[11] This presents several challenges, such as the difficulty of learning an optimal model and its policy from a partial signal, learning to cooperate or compete in non-stationary environments with distributed, simultaneously learning agents, the interplay between abstraction and influence of other agents, etc. Moreover, there are scalability, stability, and convergence issues.

Policy-based RL methods try to find the optimal policy by optimizing policy parameters with the goal to maximize the sum of total rewards. In an environment, an agent's policy is its learned behavior. Mathematically, a deterministic policy can be defined as a function mapping from a state to an action. In an uncertain environment stochastic policy can be represented with a probability distribution over actions given states. Many real world problems are stochastic, continuous, and non-stationary due to an evolving

environment or adversarial in naturewhere an environment seems to be reacting against the agent.[12]

Online learning algorithms such as Q-Learning are commonly used to address non-stationarity problems. However, Q-Learning is unable to handle complex environments where the number of states (and actions within) is large as it would require an enormous amount memory to save and update the full *state-action* table. Deep Q-Learning (DQL) is an extension to Q-Learning which uses a neural network to approximate the Q-value function. States are fed into a neural network as input values, which outputs Q-values of all possible actions. Therefore, in DQL agents continuously train on new samples in an online manner and store their learned experiences in a replay buffer that is reused as experiences from the past. As mentioned earlier, due to non-stationarity, stochasticity, and continuity of multi-agent environments, agents' policies change rapidly during training. Agents respond to this evolving environment by updating their behavior. However, this constant modification in agents' behavior causes a convergence problem. Therefore, agents' convergence to their individual optimal policy does not guarantee that the collective policies will lead to an optimal system behavior.[13-16] In other words, MARL suffers from learning stability where past experiences may not reflect the updated optimal policy, thus failing to stabilize DQL. Therefore, commonly used online learning approaches are not suitable for MARL.

### 1.3 MARL Applications in MDO

Future Army battlefields will include very complex and heterogeneous environments by connecting land, water, air, space, and cyberspace in a cohesive network. *Internet of Things (IoT)* allows collection and dissemination of digital information by deploying several devices which output a massive amount of data.[17-22] Data related technologies provide IoT solutions to help industry and government to make better decisions by exploring underlying data regularities and revealing patterns, and they provide an early warning to act upon possible threats (malware, terrorism, fraud, etc.)[23] This allows military to make agile, perceptive, resilient, and reliable decisions in a timely manner to meet mission requirements. The Military's *Multi Domain operations (MDO)* can potentially utilize IoT technology.[24-26]

MARL applications are critical to the success of MDO. There are many challenges with MARL, such as defining a suitable learning strategy for multiple RL agents, coordination by keeping track of the other learning agents for joint behavior output (violation of convergence property due to non-stationarity of agents), and scalability of algorithms to realistic problem sizes.

Joint Forces depend on the agility of MAS (robots, software, human) to maneuver in integrated operations for successful C2 (Command and Control) which becomes exponentially challenging as the number of agents grow. However, future multi-domain environments will lack sufficient training data in an extremely dynamic (non-stationary) environment. Unlike commonly used supervised learning algorithms which require massive amount of training data, agents in MARL algorithms learn their environment by interacting it and producing massive amount of data. Limited access to real-world data and an inability to meet state-of-the-art infrastructure requirement to process massive amount of information generated by RL algorithms may not only leave coalition partners vulnerable to adversarial attack, but also introduce uncertainties (data, model, platform, etc.)[21,27]

Distributed learning is commonly used in MARL to study the coordination among agents trying to solve an environmental task.[28-30] Proposed CLDE can be seen as a possible option to address scalability issue in complex military settings where mission success relies on rapidly-adaptable teams of autonomous AI systems that interact and learn from human understanding of high-level mission goals. Therefore, agility, adaptability and resiliency of these AI agents in the ever-changing battlefield scenarios is significant for warfighters to rely on them in missioncritical MDO.[31]

## 2. CENTRALIZED LEARNING

Our study focuses on the recent advances in MARL algorithms and their benefits and challenges in MAS with centralized learning based approach. *Actor-Critic* based approaches have been able to successfully address non-stationarity problem in MARL. These algorithms can be grouped based on their learning types (e.g., value or policy based). We group these algorithms based on different learning criteria, e.g., *Shared Information*, *Value Decomposition* and *Actor-Critic*. These methods are mostly unique and perform well in specific tasks, but have certain limitations.

## 2.1 Actor-Critic Methods

In Subsection 1.2 we mentioned that Q-Learning suffers from poor convergence to global optimum because it is sensitive to changes in the Q-value (action value); even a slight change in the Q-value can have an impact on an agent's policy. In MARL, *Multi-Agent Deep Deterministic Policy Gradient (MADDPG)*, *Stigmergic collaboration*, and hierarchical approaches implicitly try to find the global optimum in optimal policies, but they require enormous computing power due to high computational complexity of *Nash Equilibrium*.[32] On the other hand, while policy-based RL methods have better convergence and perform well in high-dimensional continuous space requiring less memory, they tend to converge to a local optimum and suffer from high variance. Therefore, a hybrid *actor-critic* approach, a deep reinforcement learning based approach which combines both the approaches, is used to address these challenges. Actor (policy gradient update) and critic (state-value function) make an efficient algorithm by employing the *Temporal Difference (TD)* method to make policy updates.[2]

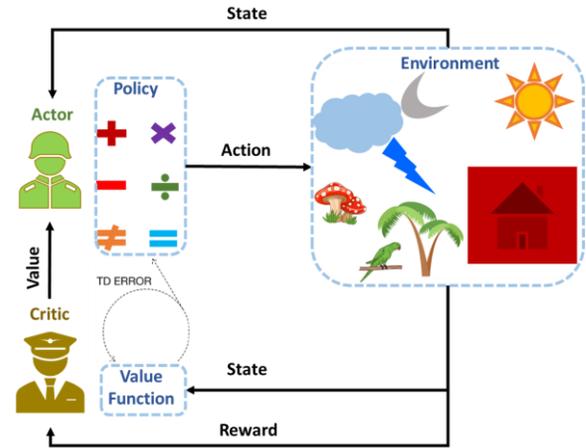

Figure 2: Illustration of actor-critic based reinforcement learning. At each step, actor interacts with the environment by taking policy-based action. Critic (estimated value function) receives a reward and transitions to a new state. After each action selection, the critic evaluates the new state to determine whether the actions taken by the actor are good or bad.

## 2.2 A2C and A3C strategies

Let (*s, a*) be a state-action pair, then an *Advantage*

function is given by $A(s, a) = Q(s, a) - V(s)$. This *Advantage* function can replace the *Value* function to address high variance policy networks and stabilize the model. While the *Value* function can help us decide whether the current state is the best available option or an action is needed to explore another state, the *Advantage* function can compute the improvement in action over the average action taken in that state. Therefore, a critic can be made to learn *Advantage* values instead of Q-values in order to make intelligent decisions in policy search algorithms. Two popular *Advantage* function-based actor-critic methods are *Advantage Actor-Critic (A2C)*[33] and *Asynchronous Advantage Actor-Critic (A3C)*.[34]

A3C consists of multiple independent agents (networks) with their own weights, and it does not essentially depend on using *experience reply* which often requires a large computer memory. Instead, it asynchronously processes a set of agents to interact with multiple copies of the environment in parallel. Policy gradient updates are done using the *Advantage* function. Such a strategy allows efficient and rapid exploration of *state-action* space while each agent (worker or network copy) is trained in parallel and asynchronously updates the global network with shared parameters. Resetting of agents' parameters and global network update is done constantly for a number of steps. Because of these joint parameter updates, agents and the global network share common information which includes *agent-agent* information. However, this periodic (asynchronous) update causes some agents to receive older version of the parameters which is not an optimal strategy.

On the other hand, A2C synchronously updates the global network. It waits until all agents have finished their training and then updates parameters in the global network. This makes it more efficient in using GPUs and performing over the large batch sizes.

### 2.3 Value and Policy Based Approaches

The mathematical framework of the most MARL methods focus on finding a Q-value and/or policy. They learn the optimal value-function or the Q-function, from which a greedy policy can be obtained. Let an agent at a given time step $t$ observe a state $s_t$ from state space $S$ and choose an action $a_t$ from action space $A$ to transition to state $s_{t+1}$ by receiving reward $r_t$. Then, for policy $\pi$, the value function for an agent starting from state $s$ and following this policy is given by

$$V^\pi = E_\pi[\sum_{t=1}^{\infty} \gamma^{t-1} r_{t+1} | s_0 = s] \quad (1)$$

where, $\gamma \in [0, 1]$ is the discount factor and $E$ is the expected value. As an extension given by Bellman, the following equation also holds from which optimal policy $\pi^*$ can be obtained:

$$V^{\pi^*}(s_t) = \max_a \sum_{s^{new}} p(s^{new}|s_t, a) \left[ r(s_t, a) + \gamma V^{\pi^*}(s^{new}) \right] \quad (2)$$

Similarly, the optimal *Q*-value is given by:

$$Q^{\pi^*}(s_t, a_t) = \sum_{s^{new}} p(s^{new}|s_t, a_t) \left[ r(s_t, a_t) + \gamma \max_{a^{new}} Q^{\pi^*}(s^{new}, a^{new}) \right] \quad (3)$$

In many real-life problems it is not feasible to find the optimal policy from the above equations. For MARL cases, a modified version of equation (3) is given by:[35]

$$Q_x^*(s, a_x | \pi_{-x}) = \sum_{a_{-x}} \pi_{-x}(a_{-x}, s) \left[ r(s, a_x, a_{-x}) + \gamma \sum_{s^{new}} P(s^{new}|s, a_x, a_{-x}) \max_{a_x^{new}} Q_x^*(s^{new}, a_x^{new}) \right] \quad (4)$$

where $\pi_{-x}(a_{-x}|s) = \prod_{i \in -x} \pi_i(a_i|s)$. Here **a** is the vector of actions for all agents, $a_{-x}$ is the vector of all agents except agent *x*. *P* is the transition probability among the states.

Alternatively, some MARL methods use policy as a parameter during learning to optimize a policy-based function. Policy gradient methods often depend on finding an approximation of the gradient. In Section 3, we explore recent MARL algorithms utilizing centralized learning with these approaches, and list them in Table 1.

### 2.4 Hybrid Approaches

Some MARL algorithms utilize a hybrid approach by combining different methods during the learning phase. For example, MAVEN uses a latent space which allows the fusion of value and policy based methods. This latent space gives a variable which is shared among agents and controlled by a hierarchical policy. Value-based agents condition their behavior on this latent variable (Subsection 3.1).

One example proposed a discrete-continuous hybrid action space-based approach, *Parametrized Deep Q- Network (P-DQN)*, addressing the limitations of classical DQN which are known to perform with either a discrete or continuous action space. P-DQN combines DQN for discrete action space and DDPG for continuous action space and extends DQN with deterministic policy for continuous actions.[36] Another example proposed two novel algorithms utilizing CLDE based approach, *Deep Multi-Agent Parameterized Q-Networks (Deep MAPQN)* and *Deep Multi-Agent Hierarchical Hybrid Q-Networks (Deep MAHHQN)*, which employ discrete-continuous hybrid action spaces in cooperative MARL settings.[37] Furthermore, an example of MARL framework presented two algorithms for adaptive packet routing in communication networks, which combines real-time Q-learning and the actor-critic (policy gradient) methods.[38]

Finally, an example shows a *Flexible Fully-decentralized Approximate Actor-critic (F2A2)* algorithm for *Decentralized Training Decentralized Executing (DTDE)* scheme. F2A2 solves *Mixed-Partially Observable Stochastic Game (POSG)* by a novel additive joint off-policy optimization objective and decentralizes via a separable primal- dual hybrid gradient descent type algorithm. Policy improvement and value evaluation are jointly optimized for each agent which can stabilize multi-agent policy learning. Algorithm achieves the guarantee on separability by introducing a general additive centralized objective function, and a joint *actor-critic* framework.[39]

### 3. SHARED INFORMATION DURING LEARNING

In this section, we discuss the recent MARL algorithmic approaches in chronological order to illuminate their improvements and limitations over previous methods. We also provide a deeper insight to *what information is being shared* by learning agents. A centralized algorithm learns something (e.g. a joint value function) that conditions on more information than is available to a single agent. This shared information is discarded before the decentralized phase. As new developments are arising in the MARL domain, this summary does not intend to be exhaustive.

### 3.1 In-Depth Exploration of MARL Algorithms

In Subsection 1.2 and Section 2, we discussed the non-stationarity challenges in MARL and how an *actor-critic* based approach can address this issue. Here, we discuss centralized learning algorithms *with* and *without* actor-critic approaches These algorithms are capable of addressing non-stationarity issue using a fully observable critic which involves the observation and actions of all agents and as a result the environment is stationary even though the other agents' policy changes.

(i) **Reinforcement Inter-Agent Learning (RIAL)** and **Differentiable Inter-Agent Learning (DIAL)** are proposed for fully cooperative, partially observable, sequential multi-agent decision-making problems, with the objective of maximizing a common discounted sum of rewards.[40] This work provides insights into swarm behavior and emergent phenomena. Both approaches are proposed in settings with centralized learning but decentralized execution, where communication between agents is not restricted during learning but is restricted over band-limited channels during execution. Both approaches are based on Deep Recurrent Q-learning Networks (DRQN) in the context of multi-agents collaboration. In RIAL, two separate Q- networks are used for the environment and communication actions, respectively, and each agent is trained in an end-to-end manner according to the Q-learning objective function. In DIAL, the communication signal and the environment actions are output from a common C-Net, and the entire multi-agent system forms a big neural network so that the entire multi-agent system is trained in an end-to-end manner. It allows a real-valued communication between agents that uses a back-propagation of the gradient via the communication channel, which favors the difficult emergence of an effective communication protocol between agents.

(ii) **Multi-Agent Deep Deterministic Policy Gradient (MADDPG)** algorithm using an actor-critic (local actors and a global critic) approach was proposed for cooperative and competitive scenarios.[41] Here, each agent uses a centralized critic and a decentralized actor. MADDPG learns *action-value* functions using centralized training and decentralized testing. Therefore, the critic can access the actions of all other agents allowing it to evaluate the effect of the joint policy on each agent's long-term reward. In order to address non-stationarity, each agent can learn an approximate model of other agents online policies. The learned policies can only use local information, so that they can be used by each agent without relying on further communication. Therefore, it learns a collection of policies for each agent, rather than just one. At each episode during the learning process, each agent draws uniformly a policy from its ensemble. Sampling a policy from an ensemble reduces the non-stationarity caused by multiple agents learning at the same time. Moreover, ensemble training of policies can also address the problem where agents can co-adept to each other and do not generalize well when deployed with agents that use unseen policies. Experiments in simulated environment show improvement over independent DDPG agents.

(iii) **Counterfactual Multi-Agent Policy Gradients (COMA)** is a *CLDE* based method which uses a single centralized critic for all the agents to estimate the Q-function and decentralised actors to optimise the agents' policies.[42] It addresses the *credit assignment* problem in multi-agent systems which earlier has been studied using *difference rewards*, an approach shaping the global reward such that agents are rewarded or penalized based on their contributions to the system's performance. These rewards are often found by estimating a reward function or using simulation, approaches which are not always trivial.[43, 44] COMA addresses this issue by computing a separate *counterfactual baseline* for each agent with centralized critic approach which computes an *advantage function* that uniquely relates to a particular agent. Here the *advantage function* is significant to perform a comparison between the *estimated return for the current joint action* and the *counterfactual baseline* that deprecates a single agent's action, while keeping the other agents' actions fixed. Moreover, in its deep neural network architecture, COMA uses a single forward pass of the actor and critic to efficiently compute the counterfactual advantage for each agent.

(iv) **Mean Field Q-Learning (MF-Q)** takes a different approach from the *CLDE* based methods (e.g., MADDPG and COMA). It employs mean field approximation over the joint action space in order to address the scalability issue that exists in the prior methods.[45] In these MARL approaches the input space of Q-value grows exponentially as the number of agents grow, causing the system to accumulate noise arising from other agents' exploratory actions. In MF-Q the pairwise interaction is approximated so that each agent is affected only by the mean effect from its neighbors. Thus, the parameters of the Q-function is independent of the number of agents as it effectively converts many-agent interactions into two-agent interactions (single agent vs. the distribution of the neighboring agents). While MF-Q is an off-policy learning approach, an on-policy learning approach, namely, *Mean Field Actor-Critic (MF-AC)*, is also proposed by slightly adjusting the model.

(v) **Multi-Actor Attention-Critic (MAAC)** was proposed which uses shared critics at training time but individual policies at test time using a specialised attention mechanism.[11] The critic is specific to each agent and has access to all other agents' embedded observations. The main idea is to use an attention mechanism in the critic that learns to selectively scale the contributions of the other agents. MAAC was compared with Counterfactual Multi-Agent Policy Gradients (COMA) which uses discrete actions and counterfactual (semi-centralized) baseline, and MADDPG which uses centralized value function and continuous actions. MAAC has better scalability as the dependency of the inputs is linear in the number of agents, rather than quadratic, and is more amenable to diverse reward and action structures than the previous methods.

(vi) **Value-Decomposition Networks (VDN)** addresses the reward sharing and inefficient policy sharing problems in the fully centralized and decentralized approaches.[6] This problem is described as *lazy learning* where one agent is able to learn useful policies, while the other agent is not because it is prevented from exploration in order to not hinder the first agent. In a cooperative MARL environment independent learning agents jointly optimize a single team reward which is accumulated over time. In such as setup, each agent can choose actions from its own action set. However, each agent suffers from a non-stationarity problem because the environment changes constantly as the learning behaviors of the other team agents change. Also, due to partial observability of the environment, agents may collect unproductive rewards from their teammates' unobserved behavior. VDN addresses these problems by learning an optimal linear value decomposition from the team value function and back-propagating the total Q gradient through deep neural networks representing the individual component value functions. Each agent implicitly learns the value function (not using agents' rewards) which depends only on each agent's local observations. Unlike other approaches, *individual agents learning directly from team reward* and *fully centralized agents*, *VDN* do not suffer from the same problems and shows much better performance across a range of more complex tasks.

(vii) **QMIX** is a hybrid method based on the *CLDE* approach. It adds a monotonicity constraint and a mixing network structure in order to make the learning stable, faster, and ultimately better in a controlled setup.[46] There is a one agent network for each agent that represents its individual value function. Each *action-value* function of the agent network is an input of the mixing network (a feed-forward network) that produces that monotonic value function. A neural network (hyper-network) produces the weights (restricted to be non-negative) for the other network enforcing a condition on the monotonicity of the network (a relationship between global and individual agent's *action-value* functions). QMIX can approximate those value functions for which an agent's best action depends on the actions of the other agents *at the same time step* more accurately than VDN, and can take advantage of the extra state information available during training.

(viii) **Weighted QMIX** is an extension to QMIX which addresses its limitations to represent joint action-value functions that are not bound by a monotonicity constraint.[47] Although, QMIX allows decentralized learning, its monotonic constraint with the mixing network causes it to fail in cases where important information about an agent's ordering over its own actions depends on other agents' actions. Therefore, it is not suitable in environments where an agent's best action depends on other agents' actions. Weighted QMIX addresses this problem by using a weighted projection scheme to assign more weight on the better joint actions, thus emphasizing them. However, on the downside, it introduces extra complexity and can perform poorly for some problems with greater challenges.

(ix) **Multi-Agent Variational Exploration (MAVEN)** is based on *CLDE*, extending the idea of deep exploration via Bootstrapped DQN to the QMIX algorithm with an aim to provide committed exploration.[7] It is a hybrid approach to leverage methods based on value and policy computation. It introduces a latent space which provides a hierarchical control over a variable shared by agents. Each agent has its own local observations and takes a local action, thus receiving a joint reward, in order to find the optimal *action-value* function. During the training each agent can access the full state and the *action-observation* of all other agents. It uses mutual information maximization between the trajectories and latent variables to learn a diverse set of such behaviours. QMIX comes with some challenges, such as its inefficacy to represent the true optimal action-value function in all cases. Also, for a fixed episode length $T$, it is proved that increasing the exploration rate decreases the probability of learning an optimal solution. MAVEN resolves the limitation of monotonicity on QMIX's exploration.

### 3.2 Shared Information

Centralized RL methods are often unable to find a global optimum. As discussed in Subsection 3.1, MARL suffers from *lazy learning*, and VDN addresses this problem by singling out certain agents based on their roles in join rewards. This class of algorithms use *Value Function Decomposition* in which successive algorithms, such as, QMIX and Weighted QMIX show improvement over the previously introduced centralized learning based methods.

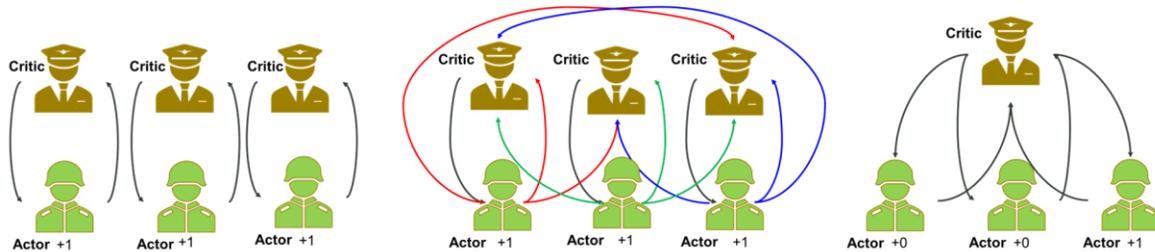

Figure 3: Illustration of some centralized learning methods using *actor-critic* approaches **(Left)** Independent learners with shared reward **(Middle)** Independent learners with shared information and shared reward **(Right)** Joint learners with independent reward.

MADDPG is a model-free approach which uses multiple critics. Each critic is able to access the observations, actions, and the target policies of all agents during learning. The role of each critic is to combine all states-actions together to use it as the input value and to obtain the corresponding Q-value by using the local reward. This causes the scalability issue. This issues can be addressed by MAAC which efficiently adjusts as the number of agents grow. On the other hand, COMA uses a single centralized critic which uses the global state, the vector of all actions, and a joint reward. This critic is shared among all agents, while the actor is trained locally for each agent with the local observation-action history. Moreover, COMA also addresses the *credit assignment* problem in multi-agent systems. MF-Q and MF-AC address these scalability issues. We provide a visual illustration of these different information sharing mechanisms in centralized learning algorithms in Figure 3 and categorize aforementioned MARL algorithms based on different criteria in Table 1.

Table 1: Categorization of MARL Methods

| Method | Shared Information | Learning Type | Algorithm Type | Reward/Policy (During Learning & Execution) | State-Action |
|---|---|---|---|---|---|
| RIAL & DIAL | Parameters & Real-Valued Message | CLDE | Value-Based | Global | Local |
| MADDPG | Joint observation and joint actions | CLDE | Actor-Critic | Local | Global & Local |
| COMA | Joint observation and joint actions | CLDE | Actor-Critic | Global | Global & Local |
| MF-Q & MF-AC | Policy gradient-based optimizers | CLDE | Value-Based & Actor-Critic | Global & Local | (Global & Global)-(Global & Local) |
| MAAC | Joint Observation and Joint Actions | CLDE | Actor-Critic | Global & Local | Global & Local |
| VDN | Joint Action-Value Function | Centralized with Value Decomposition | Value-Based | Global | Local |
| QMIX | Joint Action-Value Function | Centralized with Value Decomposition | Value-Based | Global | Global & Local |
| Weighted QMIX | Joint Action-Value Function | Centralized with Value Decomposition | Value-Based | Global | Global & Local |
| MAVEN | Action-Observation Histories & Full State | CLDE | Hybrid: Value & Policy-Based | Local | Local |

In RIAL, all agents receive parameters from a single network which is used during the learning phase. Then, decentralization is used during the execution phase. Here, agents can achieve different behavior by receiving different observations. Because it does not pass gradients between agents, no end-to-end learning across agents is used in RIAL. Nevertheless, end-to-end learning is possible from the perspective of an agent. However, RIAL is unable in avail a centralization approach completely because no agent-to-agent feedback is permitted about their communication actions.

In DIAL, centralization allows learning by passing the real-valued message between agents. This trick can restrict the communication actions between agents in a controlled fashion. Therefore, end-to-end learning is achieved across agents by permitting the gradients through a communication channel. Then, during decentralized execution, only the permitted real-valued messages are represented in discrete quantities and mapped to the discrete set of communication actions. Gradient sharing makes DIAL essentially a deep learning approach.

MADDPG, COMA, and MAAC are actor-critic based approaches where during centralized learning each agent receives joint observations and joint actions from all agents using the agents' current policies. This information is used by a centralized critic to compute the joint Q-value and for training the joint Q-value function. This implies that during learning phase each agent is able to access the joint observations and joint policy. MADDPG and MAAC belong to a class of policy-based approaches where each agent is assigned a critic during centralized learning. Each critic receives the actions of all agents plus the global state information and returns the centralized action-value for each agent. These algorithms solve the Mixed-POSG problem such that each agent only needs to optimize its own expected cumulative return. Therefore, each agent does not need to access the joint cost. On the other hand, COMA uses a single centralized critic for all the agents to estimate the Q-functions. Here, each agent has its own gradient which is computed using the shared critic during learning. The agents also share the parameters of their own policy.

MF-Q and MF-AC use the policy gradient-based optimizers during learning.

VDN, QMIX, and Weighted-QMIX learn an optimal linear value decomposition from the centralized joint action-value function. A deep neural network architecture back-propagates the total Q gradients. Each agent implicitly learns the value function (not using agents' rewards) which depends only on each agent's local observations. Therefore, decentralized execution can be carried out for each agent solely by choosing greedy actions with respect to its value (utility) function. In QMIX, the agents share a mixing function, which is discarded at the execution phase. They also share the parameters of their agent networks, which implicitly define their policies.

MAVEN shares the action-observation histories of all the agents and full state for centralized learning. It combines the value and policy-based methods in a latent space. A hierarchical policy is used to control the behavior of value-based agents by putting a condition on the shared latent variable which can be set constant for a complete episode. Therefore, each joint action-value function would behave as a joint exploratory action.

## 4. CONCLUSION

In this paper we contributed a review of recent MARL algorithms and characterized them based on information sharing between agents during learning. While the overall mechanism of these algorithms is generally understood by the research community, the information sharing mechanism during learning phase is not well-understood. Our effort intended to illuminate this learning process by focusing on *what information is being shared*. We also elaborated on the underlying mathematical framework used in the development of these algorithms and highlighted their strengths and limitations. We provided this review summary in a chronological order and addressed their utility in enhancing future *ARMY* warfighter capabilities in MDO.

Though the focus of this paper was on centralized learning in MARL, we also touched the base of decentralized learning. Future work will explore decentralized learning approaches in MARL and benchmark them against centralized learning. Moreover, we will explore how these algorithms can be used for mission critical events based on joint agent behaviors in predictive learning.

## ACKNOWLEDGMENTS

This research was supported by Oak Ridge Associated Universities (ORAU) for the U.S. DEVCOM Army Research Laboratory Grant# W911NF-16-2-0008. The views and conclusions contained in this document are those of the authors and should not be interpreted as representing the official policies, either expressed or implied, of the Army Research Laboratory or the U.S. government. The U.S. government is authorized to reproduce and distribute reprints for government purposes notwithstanding any copyright notation herein.